\documentstyle{lamuphys}
\newcommand {\cA}{{\cal A}}

\newcommand {\cD}{{\cal D}}

\newcommand {\cG}{{\cal G}}

\newcommand {\cL}{{\cal L}}

\newcommand {\cW}{{\cal W}}

\newcommand{\bR}{{\bf R}}

\newcommand{\bV}{{\bf V}}
\newcommand{\bW}{{\bf W}}

\def\a{\alpha}
\def\b{\beta}

\def\d{\delta}

\def\f{\phi}

\def\G{\Gamma}

\def\l{\lambda}

\def\o{\omega}

\def\q{\theta}
\def\r{\rho}
\def\s{\sigma}
\def\t{\tau}

\def\z{\zeta}
\def\D{\Delta}
\def\F{\Phi}
\def\J{\Psi}
\def\L{\Lambda}
\def\O{\Omega}

\newcommand{\ad}{{\dot{\alpha}}}                           
\newcommand{\bd}{{\dot{\beta}}}                            
\newcommand{\ve}{\varepsilon}                            

\newcommand{\pa}{\partial}                           
\newcommand{\hf}{\frac12}

\def\Bar#1{\overline{#1}}

\newcommand{\be}{\begin{equation}}
\newcommand{\ee}{\end{equation}}
\newcommand{\bea}{\begin{eqnarray}}
\newcommand{\eea}{\end{eqnarray}}
\newcommand{\non}{\nonumber}

\begin{document}
\setcounter{page}{0}
\thispagestyle{empty}

\begin{flushright}
hep-th/9810040
\end{flushright}

\vspace{2cm}

\begin{center}
{\bf   Covariant Harmonic Supergraphity for} \\
{\bf N = 2 Super Yang--Mills Theories}
\footnote{Based on talks given by I. Buchbinder and 
S. Kuzenko at the International Seminar ``Supersymmetries and Quantum Symmetries'',
July 1997, Dubna; to be published in the proceedings.}
\vspace{5mm}

\large{
Ioseph Buchbinder${}^1$, Sergei Kuzenko${}^2$, and Burt Ovrut${}^3$ ${}^4$
} 
\vspace{5mm}

\small{
${}^1$ Department of Theoretical Physics,
Tomsk State Pedagogical University, \\ Tomsk 634041, Russia\\
\vspace{3mm}

${}^2$ Department of Physics,
Tomsk State University, \\
Lenin Ave. 36, Tomsk 634050, Russia \\
\vspace{3mm}

${}^3$ Department of Physics, University of
Pennsylvania, \\ Philadelphia, PA 19104-6396, USA
\vspace{3mm}

${}^4$ School of Natural Sciences,
Institute for Advanced Study, \\ Olden Lane, Princeton, NJ 08540, USA
}

\end{center}
\vspace{1cm}
 
\begin{abstract}
We review the background field method for general $N=2$ super
Yang-Mills theories formulated in the $N=2$ harmonic superspace.
The covariant harmonic supergraph technique is then applied to
rigorously prove the $N=2$ non-renormalization theorem as well as to
compute the holomorphic low-energy action for the $N=2$ $SU(2)$ pure
super Yang-Mills theory and the leading non-holomorphic low-energy
correction for $N=4$ $SU(2)$ super Yang-Mills theory.
\end{abstract}

\vfill
\begin{flushleft}
August 1998
\end{flushleft}

\newpage
\title{Covariant Harmonic Supergraphity for \\
N = 2 Super Yang--Mills Theories}

\author{Ioseph\,Buchbinder\inst{1} \and Sergei\,Kuzenko\inst{2} \and
Burt\,Ovrut\inst{3}$~$ \inst{4} }

\institute{Department of Theoretical Physics,
Tomsk State Pedagogical University, \\ Tomsk 634041, Russia
\and
Department of Physics,
Tomsk State University, \\
Lenin Ave. 36, Tomsk 634050, Russia
\and
Department of Physics, University of
Pennsylvania, \\ Philadelphia, PA 19104-6396, USA
\and
School of Natural Sciences,
Institute for Advanced Study, \\ Olden Lane, Princeton, NJ 08540, USA
}
\maketitle

\begin{abstract}
We review the background field method for general $N=2$ super
Yang-Mills theories formulated in the $N=2$ harmonic superspace.
The covariant harmonic supergraph technique is then applied to
rigorously prove the $N=2$ non-renormalization theorem as well as to
compute the holomorphic low-energy action for the $N=2$ $SU(2)$ pure
super Yang-Mills theory and the leading non-holomorphic low-energy
correction for $N=4$ $SU(2)$ super Yang-Mills theory.
\end{abstract}
\section{Introduction}
Manifest covariance is one of the imperative principles in modern
theoretical physics. It means that any physical theory
possessing some symmetries must be formulated and studied in
such a form where all the symmetries are manifest both at the
classical and quantum levels.

The present paper is a brief review of recent progress in
constructing the manifestly covariant quantum formulation for
the $N=2$ supersymmetric Yang-Mills (SYM) theories
(Buchbinder et al. (1998b,c,d))
on the base of the $N=2$ harmonic
superspace developed by V.I. Ogievetsky and collaborators
(\cite{gikos}).  As we understand now, the harmonic superspace approach
is an elegant and universal setting to formulate general $N=2$ SYM
theories (\cite{gikos}) and $N=2$ supergravity
(Galperin et al. (1987a,b))
in a manifestly supersymmetric way.
Its universality follows simply from the fact that all
known $D=4$, $N=2$ supersymmetric theories can be naturally realized
in harmonic superspace. In particular, the formulations for $N=2$
SYM theories in the conventional $N=2$ superspace
(\cite{hst}) and in the $N=2$ projective superspace
(\cite{lr}) turn out to be gauge fixed and
truncated versions, respectively, of that in harmonic superspace.
It is the $N=2$ harmonic superspace which allows us to realize
the general $N=2$ SYM theories in terms of unconstrained superfields.
Therefore, just the harmonic superspace approach is an adequate and
convenient base for developing $N=2$ supersymmetric quantum field
theory.

The manifestly $N=2$ supersymmetric Feynman rules in
harmonic superspace have been developed by Galperin et al. (1985a,b).
One of the basic purposes of the present paper is to extend these
rules in order to have manifest gauge invariance along with $N=2$
supersymmetry. As is well known, the most efficient way to realize such
a goal is the background field method.

The paper is organized as follows. In section 2 we review the
(harmonic) superspace formulation for the $N=2$ SYM theories.
Section 3 is devoted to the presentation of the background field
method for such theories. In section 4 we use the background
field formulation developed to prove the $D=4$, $N=2$
non-renormalization theorem. The structure of the one-loop effective
action is discussed in section 5. Finally, in section 6 we compute
the low-energy holomorphic action for the pure $N=2$ $SU(2)$
SYM theory as well as the non-holomorphic action for the $N=4$ $SU(2)$ SYM
theory.

\section{N = 2 super Yang-Mills theories in superspace}
We start with a brief review of $N=2$ SYM theories in superspace.

\subsection{N = 2 SYM in standard superspace}
The constrained geometry of $N=2$ super Yang-Mills field is formulated
in standard $N=2$ superspace $\bR^{4|8}$
with coordinates
$z^M\equiv(x^m,\theta_i^\alpha, {\bar\theta}^i_{\dot\alpha})$
in terms of the gauge covariant derivatives
\be
{\cal D}_M\equiv ({\cal D}_m,{\cal
D}^i_\alpha,{\Bar{\cal D}}_i^{\dot\alpha})= D_M+{\rm i}\cA_M \;, \qquad
\cA_M=\cA_M^a(z)T^a
\label{covderivatives}
\ee
satisfying the algebra (\cite{gsw})
\bea
& \{\cD^i_\a,{\Bar \cD}_{{\dot\alpha j}}\}= -2{\rm i}\delta^i_j
\cD_{\a{\dot\alpha}} \;, \non \\
&  \{\cD^i_\a, \cD^j_\b \}=2{\rm i}\ve_{\a \b}
\ve^{ij}{\Bar \cW}\;,\qquad \{ {\Bar \cD}_{{\dot\a} i},
{\Bar \cD}_{{\dot\b}j}\}=2{\rm i}\ve_{{\dot\alpha}{\dot\beta}}
\ve_{ij}\cW \;.
\label{cdalgebra}
\eea
Here $D_M\equiv(\partial_m,D^i_\alpha,{\Bar D}^{\dot\alpha}_i)$ are the
flat covariant derivatives, $T^a$ are the generators of the gauge group.
The covariantly chiral strength $\cW$ satisfies the Bianchi identities
\be
{\Bar \cD}_{\ad i} \cW = 0\;, \qquad
\cD^{\a (i} \cD^{j)}_\a \cW = {\Bar \cD}^{(i}_\ad {\Bar \cD}^{j) \ad}
\Bar \cW   \;.
\ee
The covariant derivatives and a matter superfield multiplet $\varphi(z)$
transform as follows
\be
{\cal D}'_M = {\rm e}^{{\rm i}\tau}{\cal D}_M {\rm e}^{-{\rm i}\tau}\;,
\qquad \quad \varphi'= {\rm e}^{{\rm i}\tau}\varphi
\label{tautransf}
\ee
under the gauge group. Here $\tau=\tau^a(z)T^a$, and
$\tau^a={\bar\tau}^a$ are unconstrained real parameters. The set of all
transformations (\ref{tautransf}) is said to form the $\tau$-group.

The gauge invariant action of the $N=2$ pure SYM theory reads (\cite{gsw})
\be
S_{{\rm SYM}}=\frac{1}{2g^2} {\rm tr}\int {\rm d}^4x {\rm d}^4 \q \, \cW^2=
\frac{1}{2g^2}{\rm tr} \int {\rm d}^4x {\rm d}^4{\bar \q}\, {\Bar \cW}^2 \;.
\label{symaction}
\ee

\subsection{N = 2 SYM in harmonic superspace}
To realize the $N=2$ pure SYM theory as a theory of unconstrained dynamical
superfields, \cite{gikos}
extended the original superspace to $N=2$ harmonic superspace
$\bR^{4|8} \times S^2$.
A natural
global parametrization of $S^2 = SU(2)/ U(1)$ is that in terms of
the harmonic variables $({u_i}^-\,,\,{u_i}^+) \in SU(2)$
which parametrize
the automorphism group of $N=2$ supersymmetry,
\be
u^+_i = \ve_{ij}u^{+j}\;, \qquad \overline{u^{+i}} = u^-_i \;,
\qquad u^{+i}u_i^- = 1 \;.
\ee
Tensor fields over $S^2$ are in a one-to-one correspondence with
functions on $SU(2)$ possessing definite harmonic $U(1)$-charges. A function
$\J^{(p)}(u)$ is said to have the harmonic $U(1)$-charge $p$ if
$$
\J^{(p)}({\rm e}^{{\rm i}\varphi} u^+,{\rm e}^{-{\rm i}\varphi} u^-)=
{\rm e}^{{\rm i} p \varphi } \J^{(p)}(u^+,u^-)\;, \qquad
|{\rm e}^{{\rm i}\varphi}| = 1\;.
$$
A function $\J^{(p)}(z,u)$ on $\bR^{4|8}
\times S^2$ with $U(1)$-charge $p$ is called a harmonic $N=2$
superfield.

Introducing a new basis of covariant derivatives
\be
{\cal D}^\pm_\alpha={\cal
D}^i_\a u^\pm_i \;, \qquad {\Bar{\cal D}}^\pm_{\dot\alpha}= {\Bar{\cal
D}}^i_{\dot\alpha} u^\pm_i
\ee
the covariant derivative algebra (\ref{cdalgebra}) implies
\be
\{{\cal D}^+_\alpha,{\cal
D}^+_\beta\}=\{{\Bar{\cal D}}^+_ {\dot\alpha}, {\Bar{\cal
D}}^+_{\dot\beta}\}= \{{\cal D}^+_\alpha,{\Bar{\cal D}}^+_\bd\}=0
\label{gran}
\ee
and, hence,
\be
{\cD}^+_\a = {\rm e}^{-{\rm i}\O}D^+_\a {\rm e}^{{\rm i}\O}\;, \qquad
{\Bar{\cD}}^+_{\dot\a} = {\rm e}^{-{\rm i}\O}{\Bar D}^+_{\dot\a}
{\rm e}^{{\rm i}\O}
\label{bridge}
\ee
for some Lie-algebra valued harmonic superfield $\O = \O^a(z,u)T^a$ with
vanishing $U(1)$-charge, which is called the `bridge'.

As a consequence of (\ref{gran}), one can define covariantly
analytic superfields constrained by
\be
{\cD}^+_\a \Phi^{(p)}={\Bar{\cD}}^+_{\dot\alpha} \Phi^{(p)}=0\;,
\label{10}
\ee
where $\Phi^{(p)}(z,u)$ carries $U(1)$-charge $p$
and can be represented as follows
\be
\Phi^{(p)}= {\rm e}^{-{\rm i}\O}\phi^{(p)}\;, \qquad
D^+_\a \phi^{(p)}={\Bar D}^+_{\dot\a}\phi^{(p)}=0\;.
\ee
The superfield $\phi^{(p)}$ is, in general, an unconstrained function
over an analytic subspace of the harmonic superspace (\cite{gikos})
parametrized by
\be
\z \equiv\{
x^m_A,\q^{+\a},{\bar\q}^+_{\dot\a}, u^\pm_i \}\;, \quad \qquad \f^{(p)}(z,u)
\equiv  \f^{(p)} (\z)\;,
\label{analytsub}
\ee
where (\cite{gikos})
\be
x^m_A = x^m - 2{\rm i} \q^{(i}\s^m {\bar \q}^{j)}u^+_i u^-_j \;,\qquad
\q^\pm_\a = \q^i_\a  u^\pm_i \;,
\qquad {\bar \q}^\pm_{\dot\a}={\bar
\q}^i_\ad u^\pm_i\;.
\label{analbasis}
\ee
That is why such superfields are called analytic.

The analytic subspace (\ref{analytsub}) is closed under
$N=2$ supersymmetry transformations
and real with respect to the generalized conjugation
$\; \breve{} \;\equiv \;\stackrel{\ast}{\bar{}}$
(\cite{gikos}), where the operation ${}^\ast$ is defined by
\be
(u^+_i)^\ast = u^-_i\;, \qquad
(u^-_i)^\ast = - u^+_i \quad \Rightarrow \quad
(u^{\pm}_i)^{\ast \ast} = - u^{\pm}_i\;.
\ee
A remarkable property of this generalized conjugation (called below
the `smile-conjugation') is that it allows us
to consistently define
real analytic superfields with even $U(1)$-charge.

Without loss of generality, the bridge $\O$ (\ref{bridge}) can be chosen
to be real with respect to the smile-conjugation, $\breve{\O} = \O$.
The bridge possesses a richer gauge freedom than the original $\tau$-group.
Its transformation law reads
\be
{\rm e}^{{\rm i}\O'}={\rm e}^{{\rm i}\l}{\rm e}^{{\rm i}\O}
{\rm e}^{-{\rm i}\tau}
\ee
with an unconstrained analytic gauge parameter $\lambda=\lambda^a(\zeta)
T^a$ being real with respect to the smile-conjugation,
$\breve{\l}{}^a =\l^a$.
The set of all
$\lambda$-transformations form the so-called $\lambda$-group
(\cite{gikos}). The $\tau$-group acts on $\Phi^{(p)}$ and leaves
$\phi^{(p)}$ unchanged while the $\lambda$-group acts only on
$\phi^{(p)}$ as follows
\be
\phi'^{(p)}={\rm e}^{{\rm i}\lambda}\phi^{(p)}\;.
\ee
The superfields $\Phi^{(p)}$ and $\phi^{(p)}$ are said to correspond
to the $\tau$- and $\lambda$-frames respectively.

The $\l$-frame is most useful to work with the covariantly
analytic superfields.
At the same time, it is the $\l$-frame in which a single unconstrained
prepotential of the $N=2$ SYM theory naturally emerges.
Let us, first of all, introduce the harmonic derivatives
(\cite{gikos})
\be
D^{\pm\pm}=u^{\pm i}\frac{\partial}{\partial u^{\mp i}} \;, \qquad
D^0=u^{+i}\frac{\partial}{\partial u^{+i}}-u^{-i}
\frac{\partial}{\partial u^{-i}}\;,
\ee
where $D^{\pm \pm}$ are two
independent derivatives on $S^2$, and $D^0$ is
the operator of $U(1)$ charge, $D^0 \F^{(p)} = p\, \F^{(p)}$.
Operators
${\cal D}_{\underline{M}}\equiv({\cD}_M, D^{++},D^{--},
D^0) $
form a full set of gauge covariant derivatives in the $\t$-frame.
The $\lambda$-frame is defined by the following transform
\bea
&{\cal D}_{\underline{M}} \quad \rightarrow \quad
\nabla_{\underline{M}}= {\rm e}^{{\rm i}\O}{\cD}_{\underline{M}}
{\rm e}^{-{\rm i}\Omega}\;, \qquad \F^{(p)} \quad \rightarrow \quad
\f^{(p)} = {\rm e}^{{\rm i}\O} \F^{(p)}  \\
& \nabla^+_\a = D^+_\a\;, \qquad{\Bar\nabla}^+_{\dot\a}=
{\Bar D}^+_{\dot\alpha}\;, \qquad
 \nabla^{\pm\pm}
=D^{\pm\pm}+{\rm i}V^{\pm\pm}\;.
\eea
Since $[\nabla^{++}, \nabla^+_\a] = [\nabla^{++}, {\Bar \nabla}^+_\ad]=0$,
the connection $V^{++}=V^{++a}T^a$ is
a real analytic superfield, $\breve{V}^{++}=V^{++}$,
$D^+_\a V^{++}={\Bar D}^+_{\dot\a}V^{++}=0$, and its
transformation law is
\be
V'^{++}=e^{{\rm i}\lambda}V^{++}e^{-{\rm i}\lambda}
-{\rm i}\,{\rm e}^{{\rm i}\lambda}D^{++}{\rm e}^{-{\rm i}\lambda}\;.
\label{v++gaugetr}
\ee
The analytic superfield $V^{++}$ turns out to be the single unconstrained
prepotential of the
pure $N=2$ SYM theory and all other objects are expressed
in terms of it.
In particular, action (\ref{symaction}) can be rewritten via
$V^{++}$ as follows (\cite{zupnik})
\be
S_{{\rm SYM}}=\frac{1}{g^2} {\rm tr}\,
\sum\limits_{n=2}^\infty\frac{(-{\rm i})^n}
{n}\int {\rm d}^{12}z \,{\rm d}^n u
\frac{V^{++}(z,u_1)
\dots V^{++}(z,u_n)}
{(u^+_1u^+_2)\dots(u^+_nu^+_1)}\;.
\label{v++action}
\ee
The rules of integration over $SU(2)$ as well as the properties
of harmonic distributions were given by
\cite{gikos} and \cite{gios1}.

\subsection{Supersymmetric matter}
Harmonic superspace provides us with two possibilities to describe
$N=2$ supersymmetric matter in terms of unconstrained analytic
superfields (\cite{gikos}). A charge hypermultiplet, transforming
in a complex representation $R_q$ of the gauge group, is described by an
unconstrained analytic superfield $q^+(\z)$ and its conjugate
$\breve{q}^+(\z)$ ($q$-hypermultiplet). A neutral
hypermultiplet, transforming
in a real representation $R_\o$ of the gauge group, is described by an
unconstrained analytic real superfield $\o(\z)$, $\breve{\o} = \o$,
($\o$-hypermultiplet). The matter action reads
\be
S_{{\rm MAT}}=
- \int {\rm d}\z^{(-4)}\,
\breve{q}{}^+ \nabla^{++}q^+ -
\frac{1}{2}\int {\rm d}\zeta^{(-4)}
\,\nabla^{++}\omega^{\rm T}\,\nabla^{++}\omega\;,
\label{hypaction}
\ee
where the integration is carried out over the analytic subspace
(\ref{analytsub}).

\section{Background field quantization}
To quantize the pure $N=2$ SYM theory, we split $V^{++}$ into
{\it background} $V^{++}$ and {\it quantum} $v^{++}$ parts
\be
V^{++}\;\;\rightarrow \;\; V^{++}+g\;v^{++}\;.
\label{splitting}
\ee
Then, the original infinitesimal gauge transformations
(\ref{v++gaugetr}) can be
realized in two different ways:  \\
(i) {\it background transformations}
\be
\d_{\rm B} V^{++}=-D^{++}\lambda -
{\rm i}[V^{++},\lambda]=-\nabla^{++}\lambda\;,
\qquad \d_{\rm B} v^{++}={\rm i}[\lambda,v^{++}]
\label{backtr}
\ee
(ii) {\it quantum transformations}
\be
\d_{\rm Q} V^{++}=0\;, \qquad \d_{\rm Q} v^{++}=-\frac{1}{g}\nabla^{++}\l
-{\rm i}[v^{++},\lambda]\;.
\label{quantr}
\ee
It is worth pointing out that the form of the
background-quantum splitting (\ref{splitting}) and
the corresponding background and quantum transformations
(\ref{backtr}), (\ref{quantr}) are much more analogous
to the conventional Yang-Mills
theory than to the $N=1$ non-abelian SYM model.
Our aim now is to construct an effective action as a gauge-invariant
functional of the background superfield $V^{++}$.

Upon the splitting (\ref{splitting}),
the classical action (\ref{v++action}) takes the form
\be
S_{{\rm SYM}}
=S_{{\rm SYM}}[V^{++}]
+\frac{1}{4g}{\rm tr}\,\int {\rm d} \zeta^{(-4)}
\,v^{++}({\Bar D}^+)^2
{\Bar W}_\lambda
+ \Delta S\;,
\label{splittedaction}
\ee
where $\D S [v^{++},V^{++}]$ reads
\be
\Delta S = - {\rm tr}\,
\sum\limits_{n=2}^\infty {\frac{(-{\rm i}g)}{n}}^{n-2}
\int {\rm d}^{12}z\,{\rm d}^nu
\frac{v_\tau^{++}(z,u_1)
\dots v_\tau^{++}(z,u_n)}
{(u^+_1u^+_2)
\dots(u^+_nu^+_1)}\;.
\label{quantaction}
\ee
Here $W_\lambda$ and $v_\tau^{++}$
denote the $\lambda$- and
$\tau$-transforms of $W$ and $v^{++}$, respectively,
with the bridge $\Omega$ corresponding to the background covariant
derivatives constructed on the base of the
background connection $V^{++}$. The quantum action
$\Delta S$ given in (\ref{quantaction}) depends on $V^{++}$ via the
dependence of $v^{++}_\tau$ on $\Omega$, the latter being a complicated
function of $V^{++}$.
Each term in the action (\ref{splittedaction})
is manifestly invariant with respect to the background gauge transformations.
The linear in $v^{++}_\tau$ term in (\ref{splittedaction}) determines the
equations of motion. This term should be dropped when considering the
effective action.

To construct the effective action, we can use the Faddeev-Popov
Ansatz. Within the framework of the background field method,
we should fix only the quantum transformations (\ref{quantr}).
Let us introduce the gauge fixing function in the form
\be
{\cal F}^{(4)} = \nabla^{++}v^{++}\;, \qquad
\d_{\rm Q} {\cal F}^{(4)}
= \frac{1}{g}\left\{\nabla^{++}
(\nabla^{++}\lambda + {\rm i}g[v^{++},\lambda])\right\}\;.
\label{gaugefixvar}
\ee
Eq. (\ref{gaugefixvar}) leads to
the Faddeev-Popov determinant
\be
\Delta_{{\rm FP}}[v^{++},V^{++}]={\rm Det} \left\{
\nabla^{++}(\nabla^{++}+ {\rm i}gv^{++}) \right\}\;.
\label{fpdet}
\ee
To get a path-integral representation for $\Delta_{{\rm
FP}}[v^{++},V^{++}]$,
we introduce two analytic fermionic ghosts ${\bf b}$ and
${\bf c}$, in the adjoint representation of the gauge group,
and the corresponding ghost action
\be
S_{{\rm FP}}[{\bf b},{\bf c},v^{++},V^{++}]=
{\rm tr} \, \int {\rm d}\z^{(-4)}\,
{\bf b}\nabla^{++}(\nabla ^{++}{\bf c}+{\rm i}g\,[v^{++},{\bf c}])\;.
\label{ghostaction}
\ee
As a result, we arrive at the effective action
$\Gamma_{{\rm SYM}}[V^{++}]$ in the form
\bea
{\rm e}^{{\rm i}\Gamma_{{\rm SYM}}[V^{++}]}&=&
{\rm e}^{{\rm i}S_{{\rm SYM}}[V^{++}]}
\int{\cal D}v^{++}{\cal D}{\bf b}
{\cal D}{\bf c} \non \\
& \times & {\rm e}^{{\rm i}(\Delta S[v^{++},V^{++}]+
S_{{\rm FP}}[{\bf b},{\bf c},
v^{++},V^{++}])}\delta[{\cal F}^{(4)}-f^{(4)}]\;,
\label{effact1}
\eea
where $f^{(4)}(\z)$ is an external Lie-algebra valued
analytic real superfield,
and $\delta[{\cal F}^{(4)}]$
is the proper functional analytic delta-function.

To bring the effective action to a form more adapted for calculations,
we average (\ref{effact1}) with the weight
\be
\Xi[V^{++}]\,\exp\left\{\frac{{\rm i}}{2\alpha}
{\rm tr}\,\int {\rm d}^{12}z {\rm d}u_1 {\rm d}u_2\,
\,f_\tau^{(4)}(z,u_1)\frac{(u^-_1u^-_2)}{(u^+_1u^+_2)^3}f^{(4)}
_\tau(z,u_2)\right\} \;.
\label{weight}
\ee
Here $\a$ is an arbitrary (gauge) parameter, and $f^{(4)}_\t$ is
the $\t$-transform of $f^{(4)}$. The functional $\Xi[V^{++}]$
is represented as follows (\cite{bbko})
\bea
\Xi[V^{++}]&=&
\left({\rm Det}_{(4,0)}\stackrel{\frown}{\Box} \right)^{\frac{1}{2}}
\int{\cal D}\phi\,
e^{{\rm i} S_{{\rm NK}}[\phi,V^{++}]} \non \\
S_{{\rm NK}}[\phi,V^{++}]
&=& - \frac{1}{2} {\rm tr}\,\int {\rm d}\zeta^{(-4)}\,
\nabla^{++}\phi\nabla^{++}\phi
\label{40det}
\eea
with the integration variable $\phi$ being a
bosonic real analytic superfield, with its values in the Lie algebra of the
gauge group, and presenting itself a Nielsen-Kallosh ghost for the
theory. The gauge-covariant operator $\stackrel{\frown}{\Box}$ defined by
\footnote{
We use the notation
$(D^+)^4 = \frac{1}{16} (D^+)^2 ({\Bar D}^+)^2$,
$(D^\pm)^2=D^{\pm \alpha} D^\pm_\alpha$,
$({\Bar D}^\pm)^2 = {\Bar D}^\pm_{\dot{\a}}{\Bar D}^{\pm \dot{\a}}$
and similar notation for the gauge-covariant derivatives.}
\be
\stackrel{\frown}{\Box}{}=-\frac{1}{2}(\nabla^+)^4(\nabla^{--})^2
= -\frac{1}{2}(D^+)^4(\nabla^{--})^2
\label{analdal1}
\ee
moves every harmonic superfield into an analytic one, and it is equivalent
to the second-order differential operator
\bea
{\stackrel{\frown}{\Box}}{}_\t &=& {\rm e}^{-{\rm i} \O}\,
{\stackrel{\frown}{\Box}}{}\,{\rm e}^{{\rm i} \O} =
{\cal D}^m{\cal D}_m +
\frac{{\rm i}}{2}({\cal D}^{+\a}\cW){\cD}^-_\alpha+\frac{{\rm i}}{2}
({\Bar{\cD}}^+_{\dot\a}{\Bar \cW}){\Bar{\cD}}^{-{\dot\a}}
\non \\
&-&
\frac{{\rm i}}{4}({\cD}^{+\a} {\cal D}^+_\a \cW) \cD^{--}
 +\frac{{\rm i}}{8}[{\cal D}^{+\alpha},{\cal D}^-_\alpha] \cW
+ \frac{1}{2}\{{\Bar \cW},\cW \}
\label{analdal2}
\eea
when acting on the covariantly analytic superfields.
This operator is said to be the analytic d'Alambertian.
The functional ${\rm Det}_{(4,0)}\stackrel{\frown}{\Box}$,
which enters the first line of eq. (\ref{40det}),
is defined by the following path integral
\be
\left( {\rm Det}_{(4,0)} \,
{\stackrel{\frown}{\Box}}{} \right)^{-1} =
\int  {\cal D}\r^{(+4)} \cD \s
\exp \left\{ -{\rm i} \; {\rm tr} \int {\rm d}\zeta^{(-4)} \,
\r^{(+4)} {\stackrel{\frown}{\Box}}{} \,\s \right\}
\label{3a}
\ee
over unconstrained bosonic analytic real superfields $\r^{(+4)}$ and $\s$.

Upon averaging the effective action with the weight
(\ref{weight}), for $\a = -1$ one gets  the following path integral
representation (\cite{bbko})
\bea
{\rm e}^{{\rm i}\Gamma_{{\rm SYM}}[V^{++}]}&=&
{\rm e}^{{\rm i}S_{{\rm SYM}}[V^{++}]}
\left({\rm Det}_{(4,0)}\stackrel{\frown}{\Box} \right)^{\frac{1}{2}}
\non \\
& \times &
\int{\cal D}v^{++}{\cal D}{\bf b}
{\cal D}{\bf c}{\cal D}\phi\,e^{{\rm i}S_{{\rm Q}}
[v^{++},{\bf b},{\bf c},\phi,V^{++}]}\;,
\label{43}
\eea
where action $S_{{\rm Q}}$ reads
\bea
&{}& S_{{\rm Q}}[v^{++},{\bf b},{\bf c},\phi,V^{++}]=
S_2 + S_{{\rm int}} \label{44a}\\
&{}& S_2 =
{\rm tr}\,\int {\rm d}\zeta^{(-4)} \,
\left\{-\frac{1}{2}
v^{++}\stackrel{\frown}{\Box}v^{++}+
{\bf b}(\nabla^{++})^2{\bf c}
+ \frac{1}{2}\phi(\nabla^{++})^2\phi \right\}
\label{44b} \\
&{}& S_{{\rm int}}
= - {\rm tr}\,\int {\rm d}^{12}z {\rm d}^n u
\sum\limits_{n=3}^\infty
{\frac{(-{\rm i}g)}{n}}^{n-2}\frac{v^{++}_\tau(z,u_1)\dots
v^{++}_\tau(z,u_n)}{(u^+_1u^+_2)\dots(u^+_nu^+_1)} \non\\
&{}& \qquad \qquad \qquad \qquad \qquad -{\rm i}g\, {\rm tr} \,\int
{\rm d}\z^{(-4)} \nabla^{++}{\bf b}\;[v^{++}, {\bf c}]\;.
\label{44c}
\eea
Eqs. (\ref{43}--\ref{44c}) completely determine the structure of the
perturbation expansion for calculating the effective action
$\Gamma_{{\rm SYM}}[V^{++}]$ of the pure $N=2$ SYM theory
in a manifestly supersymmetric and gauge invariant form.

So far we have considered the pure $N=2$ SYM theory only.
In the general case, the classical action contains not only the
pure SYM part given by (\ref{symaction})
(or, what is equivalent, by (\ref{v++action})),
but also the matter action (\ref{hypaction}).
Our previous consideration can be easily extended to the case of the
general $N=2$ SYM theory.
The only non-trivial new information, however, is the
explicit structure of the matter superpropagators associated with
the action (\ref{hypaction}). They read as follows
\bea
\label{qgreen}
& {\rm i} &<q^+(1)\,\breve{q}^+(2)> \\
&=&  -\frac{1}{\stackrel{\frown}{\Box}{}}
{\stackrel{\longrightarrow}{(D_1^+)^4}}{}
\left\{\delta^{12}(z_1-z_2)
{1\over (u^+_1 u^+_2)^3}{\rm e}^{{\rm i}\O (1)}
{\rm e}^{-{\rm i}\O (2)} \right\}
{\stackrel{\longleftarrow}{(D_2^+)^4}}{}
\non \\
&{\rm i}& <\omega(1)\,\omega^{\rm T}(2)>
\label{ogreen} \\
&=& \frac{1}{\stackrel{\frown}{\Box}{}}
{\stackrel{\longrightarrow}{(D_1^+)^4}}{}
\left\{\delta^{12}(z_1-z_2)
{(u^-_1 u^-_2)\over (u^+_1 u^+_2)^3}
{\rm e}^{{\rm i}\O (1)}{\rm e}^{-{\rm i}\O (2)}\right\}
{\stackrel{\longleftarrow}{(D_2^+)^4}}{}\;.
\non
\eea
The Green's functions
(\ref{qgreen}) and (\ref{ogreen}) are to be used for loop calculations
in the background field approach.

The propagators of the gauge and ghost superfields follow from
(\ref{44b}). For the gauge superfield one get
\be
{\rm i}\,<v^{++}(1)\,v^{++}(2)>
=  \frac{1}{ {\stackrel{\frown}{\Box}}{}}
(D_1^+)^4
\biggl\{ \delta^{12}(z_1-z_2) \d^{(-2,2)}(u_1,u_2) \biggr\}
\label{vgreen}
\ee
with $\d^{(-2,2)}(u_1,u_2)$ being the proper harmonic delta-function
(\cite{gios1}). The propagator of the Faddeev-Popov ghosts
$\bf b$ and $\bf c$ is completely
analogous to the $\o$-hypermultiplet propagator (\ref{ogreen}).
The third ghost $\f$ contributes at the one-loop level only.

\section{The D = 4, N = 2 non-renormalization theorem}

Let us apply the covariant harmonic supergraph technique
to analyse the divergence structure of the theory. The result is
formulated as the $D=4$, $N=2$ non-renormalization theorem:
there are no ultraviolet divergences beyond the
one-loop level (\cite{hst,bko}).

Consider the loop expansion of the effective action within the
background field formulation.
Then, the effective action
is given by vacuum diagrams (that is, diagrams without external
lines) with background field dependent propagators and vertices.
In our case, the corresponding propagators are defined
by eqs. (\ref{qgreen}--\ref{vgreen}),
and the vertices can be read off from eqs.
(\ref{hypaction}) and (\ref{44c}). It is obvious that any such
diagram can be expanded in terms of background fields, and leads to a set of
conventional diagrams with an arbitrary number of external legs.

As follows from eqs. (\ref{hypaction})
and (\ref{44c}), the gauge superfield vertices
are given by integrals over the full superspace, while the matter
vertices and the Faddeev-Popov ghosts vertices are given by integrals
over the analytic subspace. Note, however, that propagators
(\ref{qgreen}--\ref{vgreen})
contain factors of $(D^+)^4$, which can be used to
transform integrals over the analytic subspace into integrals over
the full superspace if we make use of the identity
\be
\int
{\rm d}\z^{(-4)}\, (D^+)^4 \cL =  \int {\rm d}^{12}z \,{\rm d}u \, \cL\;.
\label{29}
\ee
The cost of doing this is, as a rule, the removal of one of
the two $(D^+)^4$-factors entering each matter and ghost propagator
(\ref{qgreen},\ref{ogreen}).  There is, however, one special case.
Let us consider a vertex
with two external $\o$-legs, and start to transform the corresponding
integral over the analytic subspace  into an integral over the full
superspace. To do this, we should remove the factor $(D^+)^4$
from one of the two gauge superfield propagators (\ref{vgreen}) associated
with this vertex.  As a result of transforming all integrals
over the analytic subspace into integrals over the full superspace,
each of the remaining propagators will contain, at most, one
factor of $(D^+)^4$.
Thus, any supergraph contributing to the effective action
is given in terms of the integrals over the full $N=2$ harmonic superspace.
Since this conclusion is true for each conventional supergraph in the
expansion of a given background field supergraph,
we see that an arbitrary background
field supergraph is also given by integrals over the full $N=2$
harmonic superspace. This is in complete analogy with $N=1$
supersymmetric field theories.

Once we have constructed the supergraphs with all vertices integrated
over the full $N=2$ harmonic superspace, we can perform all but one of
the integrals over the $\theta$'s, step by step and loop by loop, due to
the spinor delta-functions $\d^8(\q_i-\q_j)$ contained in the
propagators (\ref{qgreen}--\ref{vgreen}). To do this, we
remove the $(D^+)^4$-factors acting on the spinor delta-functions in the
propagators by making an integration by parts.
This allows one to obtain spinor
delta-functions without $(D^+)^4$-factors. One can then perform the
integrals over the $\q$'s. We
note that in the process of integration by parts, some of the
$(D^+)^4$-factors can act on the external legs of the supergraph.
To obtain a non-zero result in the case of
an $L$-loop supergraph, we should remove $2L$ factors of $(D^+)^4$
attached to some of the propagators using the identity
\be
\d^8(\q_1 - \q_2) (D^+_1)^4(D^+_2)^4\, \d^8(\q_1 - \q_2)
=(u^+_1u^+_2)^4 \d^8(\q_1 - \q_2) \;.
\label{30}
\ee
Thus, any supergraph
contributing to the effective action is given by a single integral over
${\rm d}^8\q$.

Now, it is not difficult to calculate the superficial
degree of divergence for the theory under consideration. Let us consider an
$L$-loop supergraph $G$ with $P$ propagators, $N_{{\rm MAT}}$ external
matter legs and an arbitrary number of gauge superfield external legs.
We denote by $N_D$ the number of spinor covariant derivatives acting on
the external legs as a result of integration by parts in the process of
transformating the contributions to a single integral over ${\rm d}^8\q$.
Then, the superficial degree of divergence $\o(G)$ of the supergraph $G$
turns out to be (\cite{bko})
\be
\o(G) =
- N_{\rm MAT} - \hf N_D \;.
\label{31}
\ee
We see immediately that all supergraphs
with external matter legs are automatically
finite. As to supergraphs with pure gauge superfield legs, they are
clearly finite only if some non-zero number of spinor covariant
derivatives acts on the external legs.
Let us now show that this is always the case
beyond one loop.

The Feynman rules for $N=2$ supersymmetric field theories in the harmonic
superspace approach have been formulated in the $\l$-frame,
where the propagators are given by eqs.
(\ref{qgreen}--\ref{vgreen}). As we have noticed, all
vertices in the background field supergraphs, including the vertices
of matter and Faddeev-Popov ghosts superfields, can be given in a form
containing integrals over the full $N=2$ harmonic superspace only. To be
more precise, this property is stipulated by the identity in
the $\l$-frame
\be
(D^+)^4 \;{\stackrel{\frown}{\Box}}{} =
{\stackrel{\frown}{\Box}}{} \;(D^+)^4 \;.
\label{32}
\ee
This identity allows one to operate with
factors $(D^+)^4$ as in case without background field, and use them to
transform the integrals over the analytic subspace into integrals over
the full superspace directly in background field supergraphs. Let us
consider the structure of the propagators in the $\l$-frame
(\ref{qgreen}--\ref{vgreen}).
The background field $V^{++}$ enters these propagators via both
${\stackrel{\frown}{\Box}}{}$ and the background bridge $\O$. The
form of the propagators
(\ref{qgreen}--\ref{vgreen})
has one drawback: if we use this form,
we can not say how many spinor derivatives act on
the external legs since the explicit
dependence of $\O$ on the background field is rather
complicated. To clarify the
situation when a number of spinor derivatives act on external legs, we use
a completely new step (in comparison with the conventional harmonic
supergraph approach developed by Galperin et al. (1985a,b)) and
transform the supergraph to the $\t$-frame (after restoring the full
superspace measure at the matter and ghost vertices).  The propagators
in the $\t$-frame are given by (\cite{bko})
\bea
{\rm i}\,<q^+_\t(1)\,
\breve{q}^+_\t(2)> &=& - \frac{1}{{\stackrel{\frown}{\Box}}{}_\t}
{\stackrel{\longrightarrow}{(\cD_1^+)^4}}{}
\left\{ \delta^{12}(z_1-z_2)
{1\over (u^+_1 u^+_2)^3} \right\}
{\stackrel{\longleftarrow}{(\cD_2^+)^4}}
\non \\
{\rm i}\,<\omega_\t(1)\,\,\omega^{\rm T}_\t(2)>
&=& \;\;\; \frac{1}{{\stackrel{\frown}{\Box}}{}_\t}
{\stackrel{\longrightarrow}{(\cD_1^+)^4}}{}
\left\{ \delta^{12}(z_1-z_2)
{(u^-_1 u^-_2)\over (u^+_1 u^+_2)^3}
\right\}
{\stackrel{\longleftarrow}{(\cD_2^+)^4}} \non \\
{\rm i}\,<v^{++}_\t(1)\,v^{++}_\t(2)>
&=&  \;\;\;\frac{1}{{\stackrel{\frown}{\Box}}{}_\t}
(\cD_1^+)^4
\biggl\{ \delta^{12}(z_1-z_2) \d^{(-2,2)}(u_1,u_2) \biggr\}\;.
\label{28}
\eea
They contain, at most, one factor of $(\cD^+)^4$
after restoring the full superspace measure
at the matter and ghost vertices.
The essential feature of these propagators is that they
contain the background field $V^{++}$ only via the
${\stackrel{\frown}{\Box}}{}_\t$ and $\cD^+$-factors; that is, only via the
$u$-independent connections $\cA_M$ (\ref{covderivatives}).
But all connections $\cA_M$ contain at least one spinor
covariant derivative acting on the background superfield $V^{++}$
(\cite{gikos}). Therefore, if we expand any background field
supergraph in the background superfield $V^{++}$, we see that each
external leg must contain at least one spinor covariant derivative.
Thus, the number $N_D$ in eq. (\ref{31})
must be greater than or equal to one.
As a consequence, $\o(G)<0$ and, hence, all supergraphs are ultravioletly
finite beyond the one-loop level. This completes the proof of the
non-renormalization theorem.

\section{The one-loop effective action}
As is clear from the above analysis, the one-loop effective action
requires a separate investigation. In what follows, we restrict
our attention to the part $\G[V^{++}]$ of effective action, which
depends on the gauge superfield only. It is $\G[V^{++}]$ which
(i) determines the one-loop ultraviolet divergences; (ii) constitutes
the effective dynamics in the Coulomb branch of $N=2$ SYM theories.

It follows from eqs. (\ref{hypaction},\ref{43},\ref{44b}) that
the one-loop effective action $\G^{(1)}[V^{++}]$ of the general
$N=2$ SYM theory reads
\bea
\G^{(1)}[V^{++}] &=&
\frac{{\rm i}}{2}\,{\rm Tr}\,{}_{(2,2)} \, \ln
{\stackrel{\frown}{\Box}}{}
- \frac{{\rm i}}{2}\,{\rm Tr}\,{}_{(4,0)} \, \ln
{\stackrel{\frown}{\Box}}{} \non \\
&-& \frac{{\rm i}}{2}\,{\rm Tr}\,{}_{ad}\, \ln
(\nabla^{++})^2 \non \\
&  +& {\rm i}\,{\rm Tr}\,{}_{R_q}\, \ln (\nabla^{++}) +
\frac{{\rm i}}{2}\,{\rm Tr}\,{}_{R_\o}\, \ln (\nabla^{++})^2 \;.
\label{one-loop}
\eea
Here the contribution in the first line
comes not only from the overal factor
in (\ref{43}), but also from the gauge superfield,
\be
\left( {\rm Det}_{(2,2)} \,
{\stackrel{\frown}{\Box}}{} \right)^{-\hf} =
\int {\cal D}v^{++} \,
\exp \left\{ -\frac{{\rm i}}{2} \; {\rm tr} \int  {\rm d}\zeta^{(-4)} \,
v^{++} {\stackrel{\frown}{\Box}}{} \,v^{++} \right\} \;.
\ee
The second line in (\ref{one-loop}) represents the joint contribution
from the Faddeev-Popov ghosts ${\bf b}$, ${\bf c}$ and the third ghost
$\f$. Finally, the third line includes the contributions from the
matter $q$- and $\o$-hypermultiplets.

The joint contribution of the Faddeev-Popov ghosts and the third ghost
differs only in sign from that of an $\o$-hypermultiplet in the
adjoint representation of the gauge group. In case of the $N=4$ SYM theory
realized in the $N=2$ harmonic superspace,
the matter sector is formed by a single
$\o$--hypermultiplet in the adjoint representation (\cite{gios2}),
and the  classical action reads
\be
S^{N=4}_{{\rm SYM}}=\frac{1}{2g^2} {\rm tr}\int {\rm d}^4x
{\rm d}^4\theta\, \cW^2
\,-\,\frac{1}{2g^2}\,{\rm tr}\,\int {\rm d}\z^{(-4)}
\,\nabla^{++}\omega\nabla^{++}\omega \;.
\label{n4}
\ee
Therefore, the corresponding one-loop effective action
is given by the first line of eq. (\ref{one-loop}),
\be
\G^{(1)}_{N=4}[V^{++}] =
\frac{{\rm i}}{2}\,{\rm Tr}\,{}_{(2,2)} \, \ln
{\stackrel{\frown}{\Box}}{}
- \frac{{\rm i}}{2}\,{\rm Tr}\,{}_{(4,0)} \, \ln
{\stackrel{\frown}{\Box}}{} \;.
\label{4}
\ee

It is the contributions in the second and third lines of (\ref{one-loop})
which (i) are responsible for all the ultraviolet divergences
of the theory and (ii) generate the low-energy holomorphic
action (see \cite{bko} for more detail). By now, we have a well elaborated
perturbative scheme to compute such quantum hypermultiplet corrections
(\cite{bbiko,bbik}). As concerns the $N=4$ SYM effective action (\ref{4}),
it is free of ultraviolet divergences, but its calculation turns out to
be a nontrivial technical problem. The point is that the one-loop
supergraphs contributing to $\G^{(1)}_{N=4}[V^{++}]$ in the harmonic
superspace approach contain coinciding harmonic singularities, that is
harmonic distributions at coinciding points. The problem of coinciding
harmonic singularities in the framework of harmonic supergraph Feynman
rules was first discussed by \cite{coinsin}.  Such singularities have
no physical origin, in contrast to ultraviolet divergences. They can
appear only at intermediate stages of calculation and should cancel
each other in the final expressions for physical quantities.
The origin of
this problem is an infinite number of internal degrees of freedom
associated with the bosonic internal coordinates.

To get rid of the one-loop coinciding harmonic singularities,
\cite{bbko} introduced,
as is generally accepted in quantum field theory,
some regularization of harmonic distributions.
Unfortunately, this regularization proved to be unsuccessful;
its use led us to the wrong conclusion  $\G^{(1)}_{N=4}[V^{++}] =0$.
In a sense, the situation in hand is similar to that with
the well-known supersymmetric regularization via dimensional reduction
which leads to obstacles at higher loops. The harmonic regularization we
used turned out to be improper already at the one-loop level.

We would like to emphasize that the problem of
coinciding harmonic singularities is associated only
with perturbative calculations of the effective action and has no
direct relation to the $N=2$ background field method itself.
The problem of coinciding harmonic singularities has been solved
by \cite{bk} for a special $N=2$ SYM background
\be
\cD^{\a (i} \cD^{j)}_\a  \cW = 0\;.
\label{onshell}
\ee
In this case the effective action can be equivalently represented in the
form
\be
\exp \left\{  {\rm i} \, \G^{(1)}_{N=4} \right\}=
\frac{
\int  {\cal D}\cG^{++}
\exp \left\{- \frac{{\rm i}}{2} \; {\rm tr} \int {\rm d}\zeta^{(-4)} \,
\cG^{++} {\stackrel{\frown}{\Box}}{} \,\cG^{++} \right\} }
{\int  {\cal D}\cG^{++}
\exp \left\{ \frac{{\rm i}}{2} \;{\rm tr} \int {\rm d}\zeta^{(-4)} \,
\cG^{++}  \,\cG^{++} \right\} }
\label{newrep}
\ee
where the analytic integration variable $\cG^{++}$ is constrained by
\be
\nabla^{++} \cG^{++} = 0\;.
\ee
Representation (\ref{newrep}) allows us to perturbatively compute
$\G^{(1)}_{N=4}$. Moreover, it can to used to prove equivalence
of the $N=2$ covariant supergraph technique to the famous $N=1$
background field formulation for the $N=4$ SYM (\cite{gsr}), when the lowest
$N=1$ superspace component of the $N=2$ vector multiplet is switched
off (\cite{bk}).

\section{Low-energy effective action}

In the Coulomb branch of the $N=2$ SYM theory, the matter hypermultiplets
are integrated out and the gauge superfield lies along a flat direction
of the $N=2$ SYM potential. If the gauge group is $SU(2)$, only
the $U(1)$ gauge symmetry survives, upon the spontaneous breakdown
of $SU(2)$, and the gauge superfield $
V^{++} = V^{++a}T^a $
($T^a =\frac{1}{\sqrt{2}} \sigma^a,~~
a =1,2,3$)
takes the form
\be
V^{++} = V^{++3} T^3 \equiv {\bV}^{++}T^3\;.
\label{67}
\ee
Here ${\bV}^{++}$ consists of two parts,
${\bV}^{++} = {\bV}_0^{++} + {\bV}_1^{++}$, where
${\bV}_0^{++}$
corresponds to a constant strength ${\bW}_0 = {\rm const}$,
and ${\bV}_1^{++}$ is an abelian gauge superfield.
The presence of ${\bV}_0^{++}$ leads to the appearance
of mass $|{\bW}_0|^2$ for matter multiplets (see \cite{bbiko}).

Since the effective action $\G[\bV^{++}]$ is gauge invariant,
it presents itself a functional of the chiral strength $\bW$ and
its conjugate $\Bar \bW$.
Assuming the validity of momentum
expansion, one can present the effective action $\G [\bW, \Bar \bW]$
in the form
\be
\G [\bW, \Bar \bW] = \left( \int {\rm d}^4 x {\rm d}^4 \q
\cL_{{\rm eff}}^{({\rm c})} + {\rm c.c.} \right)
+ \int {\rm d}^4 x{\rm d}^8 \q
\cL_{{\rm eff}}\;.
\ee
Here the chiral effective Lagrangian $\cL_{{\rm eff}}^{({\rm c})}$
is a local function of $\bW$ and its space-time derivatives,
$\cL_{{\rm eff}}^{({\rm c})} = F(\bW) + \ldots$ ,
and the higher-derivative effective Lagrangian $\cL_{{\rm eff}}$ is a
real function of $\bW$, $\Bar \bW$ and their covariant derivatives,
$\cL_{{\rm eff}} = H(\bW, \Bar \bW) + \ldots$

At the one-loop level, it is the Faddeev-Popov ghosts, the third ghost
and the matter hypermultiplets which contribute to $F(\bW)$.
As concerns the quantum correction in the first line of (\ref{one-loop}),
it contributes to the higher-derivative action $H(\bW, \Bar \bW)$.
A general analysis of covariant harmonic supergraphs given by \cite{bko}
shows that the holomorphic action $F(\bW)$ is completely generated by the
one-loop contribution. Another consequence of such an analysis is that
there is no two-loop contribution to $H(\bW, \Bar \bW)$.

The covariant harmonic supergraph technique allows us to easily compute
the holomorphic effective action. Let us restrict, for simplicity,
our consideration to the case of the pure $N=2$, $SU(2)$ SYM theory.
If we are interested in the low-energy holomorphic action,
it is proper to use the following approximation
\be
\Gamma^{(1)}_{SU(2)} [V^{++}]\approx - \Gamma_\phi[V^{++}]
\label{64}
\ee
with $\Gamma_\phi[V^{++}]$ the effective action of a real
$\omega$-hypermultiplet
in the adjoint representation of $SU(2)$ coupled to the external gauge
superfield $V^{++}$ :
\bea
e^{ {\rm i}\,\Gamma_\phi[V^{++}]} &=& \int {\cal D}\phi
\exp \left\{
- \frac{{\rm i}}{2}\,{\rm tr}\, \int  {\rm d} \zeta^{(-4)}  \,
\nabla^{++} \phi\,\nabla^{++} \phi \right\} \non \\
 \phi &=& \phi^a T^a \;, \qquad \nabla^{++}\phi = D^{++} \phi
+ {\rm i}[V^{++},\phi]\;.
\label{65}
\eea
Since the gauge superfield has the form (\ref{67}),
$\f^3$ completely decouples. Unifying $\phi^1$ and $\phi^2$
in to the complex $\omega$-hypermultiplet
$\omega = \phi^1 - {\rm i} \phi^2$,
we observe
\be
\nabla^{++}\omega = D^{++} \omega + {\rm i} \sqrt{2}{\cal V}^{++}
\omega\;,
\label{69}
\ee
hence the $U(1)$-charge of $\omega$ is $e=
\sqrt{2}$. As was shown by \cite{bbiko}, the effective actions of the
charged complex $\omega$-hypermultiplet and the charged
$q$-hypermultiplet, interacting with background $U(1)$ gauge
superfield ${\bV}^{++}$, are related by
$\Gamma_\omega [{\bV}^{++}]=2\Gamma_q [{\bV}^{++}]$ and the leading
contribution to $\Gamma_q [{\bV}^{++}]$ in the massive theory is given
by
\be
\Gamma_q [{\bV}^{++}] = \int {\rm d}^4 x {\rm d}^4 \q \;F({\bW}) +
{\rm c.c.}\;,
\qquad F({\bW})=- \frac{e^2}{64 \pi^2} {\bW}^2
\ln \frac{{\bW}^2}{\L^2}\,.
\label{70}
\ee
Here $e$ is the charge of $q^+$ (it coincides with the charge of $\omega$
in the above correspondence), $\L$ is the renormalization scale.
Since in our
case $e=\sqrt{2}$, from eqs. (\ref{64},\ref{70}) we finally obtain
the perturbative holomorphic of the $N=2$ $SU(2)$ SYM theory
\be
F_{SU(2)}^{(1)} (\bW) =\frac{1}{16 \pi^2}
\,{\bW}^2 \ln \frac{{\bW}^2}{\L^2}\;.
\label{71}
\ee
This is exactly Seiberg's low-energy effective action (\cite{seiberg})
found by integrating
the $U(1)$ global anomaly and using the component analysis.

Let us finally turn to the $N=4$ $SU(2)$ SYM theory (\ref{n4}).
Here the
non-holomorphic action $H(\bW, \Bar \bW)$ constitutes the leading
low-energy quantum correction.
Its calculation is based on the representation (\ref{newrep}).
Using the technique developed in our paper (\cite{bk}) and under
additional restrictions on the background superfields, one can
represent the effective action $\G^{(1)}_{N=4}$ by a path integral
over an unconstrained $N=1$ complex superfield $V$ and its conjugate
\bea
\exp \{  {\rm i} \, \G^{(1)}_{N=4} \}&=&
\int  \cD {\bar V}  \cD V
\exp \left\{ \frac{{\rm i}}{2} \; {\rm tr} \int {\rm d}^8 z \,
{\bar V} \Delta V \right\} \non \\
\Delta &=& \cD^a \cD_a - e W^\a \cD_\a + e {\Bar W}_\ad
{\Bar \cD}^\ad + e^2 |\f|^2 \;.
\label{del2}
\eea
Here $\f$ and $W_\a$ are the $N=1$ projections of $\bW$:
$~\f =\bW |$, $~2{\rm i} W_\a = \cD^2_\a \bW |$. Being rewritten in
terms of the $N=1$ projections, the leading non-holomorphic correction
to $\G^{(1)}_{N=4}$ takes the form
\be
\int {\rm d}^{12} z\; H(W, \Bar W)
= \int {\rm d}^{8} z \, W^\a W_\a {\Bar W}_\ad  {\Bar W}^\ad \;
\frac{\pa^4 H (\f, {\bar \f})}{\pa \f^2 \pa {\bar \f}^2} \;+\;
\cdots
\label{nh1}
\ee
To calculate $
\pa^4 H (\f, {\bar \f})/ \pa \f^2 \pa {\bar \f}^2 $, we
use a superfield proper-time technique introducing the
Schwinger kernel for the operator $\D$ (\ref{del2}). Then one gets
$\pa^4 H (\f, {\bar \f}) / \pa \f^2 \pa {\bar \f}^2 =
(4\pi \f \bar \f)^{-2}$. One can easily find a general solution to this
equation. Since the effective action of the $N=4$ SYM theory should be
scale and chiral invariant, we finally get
\be
H_{N=4}^{(1)} (\bW, {\Bar \bW}) =
\frac{1}{4(4\pi)^2 } \, \ln \frac{\bW^2}{\L^2}
\ln \frac{ {\Bar \bW}^2}{\L^2} \;.
\label{n4nonhol}
\ee
The details can be found in our work
(\cite{bk}).
This action was independently computed by
\cite{pv} and \cite{gr}.  The possibility of quantum corrections of the
form (\ref{n4nonhol}) in the effective action for the $N=4$ $SU(2)$ SYM
theory was first argued by \cite{ds}.
\vspace{5mm}

\noindent
{\bf Acknowledgements.}$~$
We are grateful to E. Buchbinder and E. Ivanov for fruitful
collaboration and numerous discussions. We are thankful to B. de Wit,
N. Dragon, J. Gates, M. Grisaru, M. Ro\v{c}ek, E. Sokatchev, S. Theisen,
and B. Zupnik for critical remarks and comments.
I. B. and S. K. acknowledge a partial support from RFBR grant,
project No 96-02-1607; RFBR-DFG grant, project No 96-02-00180;
INTAS grant, INTAS-96-0308.
B. O. acknowledges the DOE Contract No. DE-AC02-76-ER-03072
and is grateful to the Alexander von
Humboldt Foundation for partial support.

%
%

\end{document}